# A Nonlinear Master Equation for Open Quantum Systems

Roumen Tsekov

Department of Physical Chemistry, University of Sofia, 1164 Sofia, Bulgaria

A nonlinear master equation is derived, reflecting properly the entropy of open quantum systems. In contrast to linear alternatives, its equilibrium solution is exactly the canonical Gibbs density matrix. The corresponding nonlinear equation for the Wigner function accounts rigorously for the thermo-quantum entropy. It reduces at large friction to the Smoluchowski-Bohm equation in the coordinate subspace, which reflects the stochastic Langevin-Bohm dynamics. The previously derived Maxwell-Heisenberg relation for the nonequilibrium momentum dispersion of quantum Brownian particles is confirmed as well as the related quantum generalization of the classical Einstein law of Brownian motion.

The Schrödinger equation describes rigorously isolated quantum systems. It can be mathematically transformed to the Liouville-von Neumann equation, which provides alternative description in terms of the more general density operator formalism. Dividing an isolated system to subsystem and environment and integrating the Liouville-von Neumann equation over the environmental variables yield the master equation for the open quantum subsystem. It is a powerful theoretical tool for solving many problems from statistical mechanics and nonequilibrium thermodynamics. The formal Nakajima-Zwanzig equation is the most general master equation, which reduces further to the Born-Markov equation in the case of weak subsystem-environment interactions and negligible memory effects. If additionally, the complete positivity of the density matrix is required, one arrives to the Lindblad equation. All these equations are fundamentally linear[1] but thermodynamic arguments point out that the exact master equation must be nonlinear.[2,3] Indeed, the Schrödinger equation is linear for the wave function, while the classical Markov diffusion is linear for the probability density, being the square of the wave function.

In classical physics, the diffusive Markov processes obey the linear Fokker-Planck equation. A particular example, governing thermodynamic relaxation, is the Klein-Kramers equation

$$\partial_t f + \partial_p H \cdot \partial_x f - \partial_x H \cdot \partial_p f = b \partial_p \cdot (f \partial_p H + k_B T \partial_p f) \qquad (1)$$

which describes the evolution of the phase space probability density $f(p,x,t)$ of an open system of N particles, where $p$ and $x$ are 3N-dimensional vectors of all momenta and coordinates, respectively. For simplicity, the friction coefficient $b$ is considered constant for all particles but in structured environment as solids the friction depends on the positions of the subsystem particles as well.[4] Once it is annulled, Eq. (1) reduces to the Liouville equation, being equivalent to classical mechanics of the isolated subsystem. Furthermore, the special relativity is also described by Eq. (1) via the relevant Einstein expression for the Hamilton function $H(p,x)$. The relaxation term on the right-hand side drives the irreversible evolution towards thermodynamic equilibrium. The equilibrium solution is the well-known canonical Gibbs distribution $f_{eq} = \exp(-\beta H)/Z$, where $\beta \equiv 1/k_B T$ is the reciprocal temperature. The partition function $Z(N,T,V)$ determines the equilibrium free energy $F_{eq} \equiv -k_B T \ln Z = H + k_B T \ln f_{eq}$, which is the characteristic potential of the subsystem and contains the entire thermodynamic information. Thus, any problem in classical statistical mechanics and thermodynamics could be solved via Eq. (1), in principle, once the mechanical definition is specified by $H$.

It is possible to quantize Eq. (1) directly by replacing the canonical derivatives and functional products via commutators $[,]$ and anti-commutators $\{,\}$, respectively. In this way the Klein-Kramers equation transforms to the Caldeira-Leggett equation[5] for the density matrix $\hat{\rho}$ of the N-particles subsystem, which reduces to the Liouville-von Neumann equation at $b=0$,

$$\partial_t \hat{\rho} - [\hat{H},\hat{\rho}]/i\hbar = b[\hat{x},\{\hat{\rho},[\hat{x},\hat{H}]/i\hbar\}/2 + k_B T[\hat{x},\hat{\rho}]/i\hbar]/i\hbar \tag{2}$$

Conventionally, the superscript, as in the Hamiltonian $\hat{H}$, denotes quantum mechanical operators in the Heisenberg picture. It is well known that Eq. (2) is correct only at high temperature[6] and that is why its equilibrium solution differs from the rigorous quantum canonical Gibbs density operator

$$\hat{\rho}_{eq} = \exp(-\beta \hat{H})/Z \tag{3}$$

Introducing the Wigner function $W(p,x,t)$, which is the quantum analog of the classical phase space probability density $f$, Eq. (2) can be straightforward transformed to

$$\partial_t W - 2H \sin \vec{\Lambda} W / \hbar = b \partial_p \cdot (W \cos \vec{\Lambda} \partial_p H + k_B T \partial_p W) \tag{4}$$

The arrows in the super operator $\vec{\Lambda} \equiv \hbar(\vec{\partial}_x \cdot \vec{\partial}_p - \vec{\partial}_p \cdot \vec{\partial}_x)/2$ indicate the direction of differentiation and the commutators and anti-commutators change to $2i \sin \vec{\Lambda}$ and $2 \cos \vec{\Lambda}$, respectively.[7] Since Eq. (4) reduces to the Wigner-Moyal equation in the case $b=0$, it accounts rigorously for quantum mechanics on the left-hand side but the last diffusional term on the right-hand side is purely classical. This semiclassical discrepancy results in an approximate equilibrium solution. For instance, Eq. (4) reduces exactly to the classical Eq. (1) in the case of harmonic oscillators with the Hamilton function $H \equiv p^2/2m + m\omega_0^2 x^2/2$. Thus, any initial quantum correlation will disappear during the irreversible evolution and the quantum oscillators will become classical at equilibrium. Traditionally, this problem is fixed by replacing the thermal energy $k_B T$ via the mean energy $\varepsilon = (\hbar\omega_0 / 2)\coth(\beta\hbar\omega_0 / 2)$ of the quantum Brownian oscillator at equilibrium to obtain

$$\partial_t W + p \cdot \partial_x W / m - m\omega_0^2 x \cdot \partial_p W = b\partial_p \cdot [pW / m + (\hbar\omega_0 / 2)\coth(\beta\hbar\omega_0 / 2)\partial_p W] \tag{5}$$

Such an approach is, however, neither rigorous nor universal and demonstrates again the thermodynamic limitations of the Caldeira-Leggett equation (2). The enhancement of the latter to the Lindblad form fails also to reproduce Eq. (3) in general, except for harmonic oscillators.[8]

The main goal of the present paper is to improve the Caldeira-Leggett equation. In general, Markov processes do not exist either in classical or in quantum mechanics, but they are the most reliable idealizations for any dynamics. As was mentioned, the rigorous approach requires integration of the exact Liouville-von Neumann equation, which is possible only for harmonic oscillators, because of linearity of the corresponding dynamic equations. In this way the Markov Caldeira-Leggett equation is derived by employing some additional hypotheses, e.g. the factorization of the initial quantum state in the Feynman-Vernon approach. For this reason, we will try also to map the quantum dynamics on the Markov one. Due to degeneracy of our procedure for quantization, however, our master equation is postulated, not derived. A number of experiments on different systems must be conducted to find out where this equation applies. According to the Onsager nonequilibrium thermodynamics, the flow is proportional to the gradient of the relevant thermodynamic potential, which is the nonequilibrium local free energy $F \equiv H + k_B T \ln f$ now. Respecting this deeper physics, one should rewrite Eq. (1) in the form

$$\partial_t f + \partial_p H \cdot \partial_x f - \partial_x H \cdot \partial_p f = b\partial_p \cdot (f\partial_p F) \tag{6}$$

Quantizing Eq. (6) we propose the following form of the master equation

$$\partial_t \hat{\rho} - [\hat{H}, \hat{\rho}] / i\hbar = b[\hat{x}, \{\hat{\rho}, [\hat{x}, \hat{H} + k_B T \ln \hat{\rho}] / i\hbar\} / 2] / i\hbar \tag{7}$$

and it is obvious that Eq. (3) is the equilibrium solution of Eq. (7). A fundamental difference between the new master equation and Eq. (2) is the Boltzmann logarithm originating from the subsystem entropy. The classical Eq. (6) is linear due to the differentiation of the entropy, while Eq. (7) remains nonlinear owing to the noncommutative quantum algebra.[2] It is known that the exact von Neumann entropy $S \equiv -k_B tr(\hat{\rho}\ln\hat{\rho}) \neq -k_B \int W \ln W dp dx$ differs from the Shannon-Wigner entropy, which is driving the diffusion in Eq. (4), although the energy $E \equiv tr(\hat{\rho}\hat{H}) = \int HW dp dx$ is the same in both representations. The nonlinearity of Eq. (7) changes dramatically the quantum evolution of open systems by repealing the superposition principle. This requires a critical reassessment of the quantum decoherence, described traditionally via linear master equations.[9]

To demonstrate the correctness of Eq. (7), one can linearize it around the exact equilibrium density operator $\hat{\rho}_{eq} = \exp(-\beta \hat{H})/Z$ to obtain

$$\partial_t \hat{\rho} - [\hat{H}, \hat{\rho}]/i\hbar = bk_B T [\hat{x}, \{\exp(-\beta \hat{H}), [\hat{x}, \{\exp(\beta \hat{H}), \hat{\rho}\}/2]/i\hbar\}/2]/i\hbar \qquad (8)$$

The equilibrium solution of this equation is naturally Eq. (3). If one considers further the high temperature limit and linearizes the exponential operators as well, it reduces to the Caldeira-Leggett equation (2), as expected. An advantage of the linearity of Eq. (8) is that it can be directly transformed in the Wigner phase space

$$\partial_t W - 2H \sin \vec{\Lambda} W / \hbar = bk_B T \partial_p \cdot \{\exp(-\beta H \cos \vec{\Lambda}) \partial_p [\exp(\beta H \cos \vec{\Lambda})W]\} \qquad (9)$$

As is seem, the formal equilibrium solution $W_{eq} = \exp(-\beta H \cos \vec{\Lambda})/Z$ obeys the Bloch-Wigner equation $\partial_\beta (W_{eq} Z) = -H \cos \vec{\Lambda} W_{eq} Z$, as required. In the simplest case of an ideal gas, the Hamilton function $H \equiv p^2/2m$ depends on the momenta of the subsystem particles only and Eq. (4) coincides with the classical Eq. (1). Surprisingly, Eq. (9) reduces also to Eq. (1), which shows that quantum effects for free Brownian particles must be nonlinear. For harmonic oscillators the super operator $H \cos \vec{\Lambda} = H - H\vec{\Lambda}^2/2$ splits to two parts, depending on $p$ and $x$, respectively.[7] The contributions of the $x$-part cancel in Eq. (9), since it commutes with $\partial_p$. Because the second derivative on $\beta$ of the relaxation operator for Brownian harmonic oscillators equals to the operator itself multiplied by $(\hbar \omega_0/2)^2$, the latter is a linear combination of the hyperbolic sine and cosine functions of $\beta \hbar \omega_0/2$. Therefore, Eq. (9) acquires the following particular form

$$\partial_t W + p \cdot \partial_x W/m - m\omega_0^2 x \cdot \partial_p W = bk_B T \partial_p \cdot [2\sinh(\beta \hbar \omega_0/2) pW/m\hbar \omega_0 + \cosh(\beta \hbar \omega_0/2) \partial_p W] \qquad (10)$$

Both Eq. (5) and Eq. (10) are linear and possess the same exact equilibrium solution but $W_{eq}$ is derived from Eq. (10) and presumed in Eq. (5). The quantum effect in Eq. (5) is solely prescribed to diffusion, while in Eq. (10) both the friction and diffusion are quantum. The emergent friction coefficient $B \equiv b \sinh(\beta \hbar \omega_0/2)/(\beta \hbar \omega_0/2)$ agrees with the Wigner quantum transition state theory at zero barrier,[1] since $\sinh(\beta \hbar \omega_0/2)$ is inversely proportional to the partition function of a quantum oscillator. The momentum diffusion coefficient $D_p \equiv bk_B T \cosh(\beta \hbar \omega_0/2)$ is also amplified but obeys the quantum fluctuation dissipation theorem $D_p = B(\hbar \omega_0/2) \coth(\beta \hbar \omega_0/2)$. The adiabatic friction coefficient $\beta D_p = b\cosh(\beta\hbar\omega_0/2)$ is larger than the isothermal $B$ and they are interrelated via the Gibbs-Helmholtz equation $\partial_\beta(\beta B)_b = \beta D_p$. At zero temperature, the friction coefficients diverge, because $\hbar\omega_0/2$ plays the role of activation energy as well, and the harmonic oscillator drops at once in the equilibrium ground state with $W_{eq} = \exp(-2H/\hbar\omega_0)/Z$. The quantum oscillator moves in the grounds state without microscopic friction due to the tunneling effect, but to move macroscopically it should be exited first. At zero temperature the environment cannot supply the necessary excitation energy $\hbar\omega_0$, which reflects in the infinite emergent friction coefficient $B$. This effect weakens, however, by a decrease of the collision frequency $b/m$, which at zero temperature is solely due to the quantum motion of the subsystem particles in the ground state.[10]

Formally, it is possible to convert Eq. (7) in the Wigner representation

$$\partial_t W - 2H\sin\vec{\Lambda}W/\hbar = b\partial_p \cdot \{W\partial_p[\cos\vec{\Lambda}H + k_B T \ln(\cos\vec{\Lambda}W)]\} \tag{11}$$

Using the operator equality $\cos\vec{\Lambda}\exp(-\beta H\cos\vec{\Lambda}) = \exp(-\beta\cos\vec{\Lambda}H)\cos\vec{\Lambda}$ one can prove that the equilibrium solution of Eq. (11) is the exact $W_{eq} = \exp(-\beta H\cos\vec{\Lambda})/Z$ again. Extracting the Shannon-Wigner entropy, Eq. (11) can be further presented in the form of Eq. (4)

$$\partial_t W - 2H\sin\vec{\Lambda}W/\hbar = b\partial_p \cdot [W\cos\vec{\Lambda}\partial_p H + k_B T\partial_p W + k_B TW\partial_p \ln(\cos\vec{\Lambda}W/W)] \tag{12}$$

It is evident now that the last nonlinear term represents the quantum entropy, vanishing naturally in the classical limit $\hbar \to 0$. It persists even at zero temperature to ensure the correct quantum distribution in the ground state. Solving the nonlinear Eq. (12) in general is a mathematical problem more difficult than quantum mechanics of closed systems, because the Liouville-von Neuman part is much simpler than the relaxation one. However, taking the leading quantum corrections, $\sin\vec{\Lambda} \approx \vec{\Lambda} - \vec{\Lambda}^3/6$ and $\cos\vec{\Lambda} \approx 1 - \vec{\Lambda}^2/2$, and expanding the logarithm in series as well yield a semiclassical Klein-Kramers equation

$$\partial_t W + \partial_p H \cdot \partial_x W - \partial_x H \cdot \partial_p W + H\vec{\Lambda}^3 W/3\hbar = b\partial_p \cdot [W\partial_p H + k_B T\partial_p W - k_B TW\partial_p(\vec{\Lambda}^2 W/2W)] \tag{13}$$

The linear quantum term on the left-hand side is well known and vanishes for free particles and oscillators. The quantum term on the right-hand side is nonlinear and accounts for the Fisher entropy via the nonlinear Bohm quantum potential, represented in the Wigner phase space.[3] The latter originates obviously from the quantum entropy and deserves its reference as information potential. For numerical applications in chemistry, for instance, a TDDFT image of Eq. (13) is already proposed via a nonlinear dissipative thermo-quantum Kohn-Sham equation.[3]

Let us return back to the harmonic oscillators. Although the corresponding Eq. (13) is nonlinear, its solution is a normal distribution. Using a bivariate Gaussian Wigner function for every oscillator, the nonlinear quantum term acquires the linear form $k_B T \hbar^2 \partial_p W / 4(\sigma_x^2 \sigma_p^2 - \sigma_{xp}^2)$. The bath effect is significant at large friction constant $b$, where the Brownian motion of the subsystem particles becomes overdamped. In this case, the fast thermalization in the momentum subspace is already over and the observation follows solely the slow relaxation in the coordinate subspace. Because the nonlinear term is a quantum correction, one should employ therein the relevant classical expressions for the momentum dispersion $\sigma_p^2 = m k_B T$ and correlation $\sigma_{xp} = 0$ at equilibrium. Hence, substituting the Bohmian term $\hbar^2 \partial_p W / 4m\sigma_x^2$ back in Eq. (13) yields an emergent Fokker-Planck equation

$$\partial_t W + p \cdot \partial_x W / m - m\omega_0^2 x \cdot \partial_p W = b \partial_p \cdot [pW / m + (k_B T + \hbar^2 / 4m\sigma_x^2) \partial_p W] \tag{14}$$

One can see immediately that the quantum entropy increases the thermal energy by the Heisenberg momentum uncertainty, i.e. the classical environment monitors continuously the quantum subsystem by measurements. This nonequilibrium thermo-quantum Maxwell-Heisenberg relation[11] substitutes the equilibrium momentum dispersion in Eq. (5). Combining the equilibrium Maxwell-Heisenberg relation $\sigma_p^2 = m k_B T + \hbar^2 / 4\sigma_x^2$, minimal for Gaussian distributions, with the virial theorem $m\omega_0^2 \sigma_x^2 = \sigma_p^2 / m$ yields the mean energy $\varepsilon = (k_B T / 2)[\sqrt{1 + (\beta \hbar \omega_0)^2} + 1]$,[3] which is slightly higher than the exact $\varepsilon = (\hbar \omega_0 / 2) \coth(\beta \hbar \omega_0 / 2)$ due to the semiclassical approximations in Eq. (13). Both expressions coincide, however, at zero and infinite temperature. Following the standard procedure at large $b$, one can derive from Eq. (14) the Smoluchowski-Bohm equation, governing the probability density $\rho(x,t) = \int W dp$ in the coordinate subspace, which corresponds to the diagonal elements of the density matrix,

$$\partial_t \rho = \partial_x \cdot [m\omega_0^2 x \rho + (k_B T + \hbar^2 / 4m\sigma_x^2) \partial_x \rho] / b = \partial_x \cdot [\rho \partial_x (U + Q) / b + D \partial_x \rho] \tag{15}$$

The last general form is already derived via dissipative quantum hydrodynamics.[3,11] It is valid for arbitrary interaction potential $U(x)$, $D \equiv k_B T / b$ is the classical Einstein diffusion constant, the nonlinear Bohm quantum potential $Q \equiv -\hbar^2 \partial_x^2 \sqrt{\rho} / 2m\sqrt{\rho}$ is represented in the coordinate subspace, and the underling stochastic dynamics obeys the density-functional Langevin-Bohm equation. According to Eq. (15) at $T = 0$, the position dispersion $\sigma_x^2 = (\hbar / 2m\omega_0)\sqrt{1 - \exp(-4m\omega_0^2 t / b)}$ relaxes quicker than the classically-like prediction of Eq. (5) $\sigma_x^2 = (\hbar/2m\omega_0)[1-\exp(-2m\omega_0^2 t/b)]$. Both expressions tend, however, to the exact equilibrium dispersion of the ground state. The Smoluchowski-Bohm equation does not provide the exact equilibrium solution in general, due to the semiclassical approximations in the quantum entropy. In fact, the thermal fluctuations are accounted twice in the stochastic Langevin-Bohm equation, because of the temperature dependence of the quantum potential via $\rho$. It is possible, however, to improve Eq. (15) by replacing $Q$ with the corresponding free energy $k_B T \int Q d\beta$ from the Gibbs-Helmholtz relation to obtain[3,11]

$$\partial_t \rho = D\partial_x \cdot [\rho \partial_x \int_0^\beta (U+Q)d\beta + \partial_x \rho] = D\partial_x \cdot [\rho \partial_x \int_0^\beta \frac{1}{\sqrt{\rho}}(\hat{H}+2\partial_\beta)\sqrt{\rho}d\beta] \qquad (16)$$

The last integral represents the nonequilibrium free energy functional in the coordinate subspace divided by $k_B T$. Since it becomes $-\ln Z$ at equilibrium, the equilibrium distribution obeys a Bloch type equation $2\partial_\beta \sqrt{\rho_{eq} Z} = -\hat{H}\sqrt{\rho_{eq} Z}$. Solutions of the latter are the quantum canonical Gibbs distributions $\rho_n(x) = \exp(-\beta E_n)\varphi_n^2(x)/Z$, where $E_n$ and $\varphi_n$ are the eigenvalues and normalized eigenfunctions of the subsystem Hamiltonian, respectively. The well-known expression for the quantum partition function $Z = \sum \exp(-\beta E_n)$ follows from normalization. The integral in Eq. (16) is going beyond the Maxwell-Heisenberg relation and complicates additionally the relaxation of open quantum systems but one can be sure that the final equilibrium state is exact.

Finally, let us reconsider the most interesting case of an ideal gas by setting $\omega_0 \equiv 0$ above. As discussed before, in this case Eq. (5) becomes classical. The Maxwell-Heisenberg relation provides now the exact value at equilibrium, because $\sigma_x^2$ diverges in time. For free particles, Eq. (15) reduces to diffusion equation with a dispersion-dependent diffusion coefficient $D(1+\lambda_T^2/\sigma_x^2)$, where the Planck constant scales to the thermal de Broglie wave length $\lambda_T \equiv \hbar/2\sqrt{mk_B T}$. The direct integration of the standard diffusional equation $\partial_t \sigma_x^2 = 2D(1+\lambda_T^2/\sigma_x^2)$ confirms our quantum generalization of the classical Einstein law of Brownian motion[3,11]

$$\sigma_x^2 - \lambda_T^2 \ln(1+\sigma_x^2/\lambda_T^2) = 2Dt \qquad (17)$$

The quantum characteristic time $\tau \equiv \lambda_T^2/2D = b/2m\omega_2^2$ corresponds to an oscillator with the second Matsubara frequency $\omega_2 \equiv 2k_B T/\hbar$. The classical Einstein law $\sigma_x^2 = 2Dt$ holds when $t > \tau$, which is easily achieved at high temperature, but at short time a purely quantum expression $\sigma_x^2 = \hbar\sqrt{t/mb}$ follows from Eq. (17). This sub-diffusive quantum law is our central invention, being always valid at low temperature, where the quantum entropy dominates over the classical one.[11] Because Eq. (17) is derived for large time, the necessary condition to be able to see quantum effects is $2b\tau/m > 1$, i.e. the Nelson diffusion constant $\hbar/2m > D$ must be larger than the Einstein one, which is typical for light particles at low temperature and high friction.

The Planck constant appears in the present paper solely from the subsystem quantum operators. Therefore, the considered thermal bath is classical and affects the subsystem particles only via the friction constant $b$ and temperature $T$. For this reason, the Smoluchowski-Bohm equation describes classical diffusion in the fields of classical and quantum potentials. In general, the environment can be quantum as well, which complicates additionally the theoretical analysis via a time-temperature operator[10] $k_B \hat{T} \equiv (\hat{E}/2)\coth(\beta\hat{E}/2)$, where $\hat{E} \equiv i\hbar\partial_t$ is the energy operator in quantum mechanics, and more complex quantum friction,[12] which can affect the equilibrium distribution as well.[1] Therefore, it is essential to distinguish our Brownian motion of quantum particles in a classical environment from the more complex Brownian motion in a quantum environment. It is well known that $\sigma_x^2$ grows logarithmically in time for the quantum Brownian motion in an environment with non-Markov retardation at zero temperature.[13] Interestingly, this quantum bath effect can also be accounted via the thermo-quantum Maxwell-Heisenberg relation $\sigma_p^2 = mk_BT + m\hbar/t + \hbar^2/4\sigma_x^2$, enhanced by the Heisenberg time-energy uncertainty. Here, the quantum corrections are solely the first two terms in an infinite series on the powers of the Planck constant $\hbar$. The linear term accounts for the quantum environment, since it supplies pure energy, while the particle quantum contribution is given by the quadratic term, because it goes through the particle momentum. Knowing the influence of potentials in quantum mechanics, we expect a dramatic quantum effect of the position dependent friction coefficient in structured media[4] and of the nonlinear friction,[14] going beyond the Onsager linear force-flux relation.